\documentclass[preprint,12pt]{elsarticle}

\usepackage{amssymb}
\usepackage{amsmath}
\usepackage{algorithm}
\usepackage{algorithmic}
\usepackage{multirow}
\usepackage{tabularx}
\usepackage{color}
\usepackage{ulem}
\usepackage{makecell}
\usepackage[hyphens]{url}
\usepackage{breakurl}
\newtheorem{definition}{Definition}

\journal{Nuclear Physics B}

\begin{document}

\begin{frontmatter}


\title{Hiding the Trees in the Forest: Building Network Covert Channels with Hash-Based Covert Carrier Filtering}


\author[addr1]{Zexiao Zou}
\ead{20232002@mail.besti.edu.cn}

\author[addr1]{Zhiqiang Wang\corref{cor1}}
\ead{wangzq@besti.edu.cn}

\author[addr2,addr3]{Baoxu Liu}
\ead{liubaoxu@iie.ac.cn}

\author[addr1]{Yuyang Han}
\ead{20232005@mail.besti.edu.cn}

\author[addr1]{Yan Zhang}
\ead{zhangyan@besti.edu.cn}

\cortext[cor1]{Corresponding author}

\address[addr1]{Beijing Electronic Science and Technology Institute, Beijing 100070, China}
\address[addr2]{Institute of Information Engineering, Chinese Academy of Sciences, Beijing 100085, China}
\address[addr3]{School of Cyber Security, University of Chinese Academy of Sciences, Beijing 100049, China}

\begin{abstract}

As an effective anti-censorship mechanism, network covert channels can provide data privacy protection and ensure communication security. However, the covertness of existing network covert channels primarily depends on the secrecy of their covert algorithms. With the increasing depth of research in this field, the difficulty of breaking such algorithms has gradually decreased. Once the algorithm is exposed, the network covert channel can be easily detected by adversaries. To address this issue, this paper proposes a covert carrier filtering strategy based on the hash. In this strategy, a key-dependent filtering rule is introduced during the construction of the network covert channel, enabling the communicating parties to randomly and dynamically filter a sparse subset from the carrier set as the covert carrier set. This strategy not only enhances the randomness of carrier selection but also tightly couples the covertness of the network covert channel with the security of the key. We employ machine learning-based traffic analysis methods to experimentally validate the strategy in two types of network covert channels: network storage and timing covert channels. The experimental results demonstrate that the proposed strategy significantly improves the detection resistance of network covert channels. When the filter key size exceeds six bits, the impact on the detection effect of the classifier becomes quite significant. Furthermore, the processing delay for a single packet is less than 8 $\mu s$, indicating the feasibility of deploying the proposed strategy in high-speed network environments.

\end{abstract}




\end{frontmatter}



\section{Introduction}
\subsection{Background}
With the continuous advancement of the information society, cybersecurity has become a focal point of global concern. Data privacy and communication security within network environments have emerged as critical challenges requiring urgent solutions. Against this backdrop, network covert channels, as an anti-censorship mechanism designed to evade monitoring and interception, have attracted significant attention. As a branch of information hiding technology, network covert channels encode secret data by manipulating the characteristic patterns of legitimate network packets. This technique embeds confidential information within normal communication traffic, making it difficult to detect and thereby enabling the covert transmission of data. The applications of network covert channels are extensive; they can be utilized for confidential communication but may also be maliciously exploited to bypass cybersecurity defenses. The channel's covertness is a core criterion for evaluating its performance.

The covertness of a network covert channel refers to its ability to avoid discovery, including imperceptibility and undetectability. Imperceptibility, means the channel's existence remains hidden from regular users and does not interfere with normal services, making it indistinguishable at the operational level. Undetectability means the statistical features of the covert carrier should not exhibit significant alterations. The level of covertness is related to the complexity of the covert algorithm, which includes the steps for establishing the network covert channel and the methods for encoding and decoding the covert data. Current research into network covert channels has matured the exploration of their construction methods and carrier types. Wendzel et al. \cite{paper_1} used the Pattern Language Markup Language (PLML) method to categorize 109 covert channels developed between 1987 and 2013 into 11 distinct patterns, with most covert channels falling into four primary categories. Similarly, Li et al. \cite{paper_2} summarized the potential forms and construction methods of storage and timing covert channels by analyzing them from three aspects: symbol design, information encoding, and channel optimization.
Beyond the intrinsic complexity of the covert algorithm, the covertness of existing network covert channels primarily depends on the secrecy of the algorithm itself. However, ongoing research and exploration have made network covert channels increasingly susceptible to enumeration and discovery. Consequently, relying solely on the secrecy of the covert algorithm is no longer sufficient to ensure the covertness of these channels. Once the covert algorithm is exposed, an adversary can easily decipher the channel's construction method and identify the covert carrier. It should be noted that while encrypting covert data protects the information content after interception, encryption alone does not directly address the detectability of communication behavior. As noted by Howes et al. \cite{9833752} in their formalization of application-based covert channels, even when secure channels are used to encrypt communication, covert channels may still be detected due to the unique or unusual traffic patterns they induce.

\subsection{Contribution}

The covertness of traditional network covert channels relies heavily on the secrecy of the covert algorithm. Their detection can essentially be reduced to an adversary analyzing traffic to identify consistent anomalies generated by a fixed algorithm that deviate from normal patterns. Once such patterns are recognized, the network covert channel is effectively exposed. To fundamentally alter this adversarial dynamic, we introduce a key-based covert carrier filtering strategy. The core idea of this strategy is to decompose a single, global modification pattern into numerous sparse, pseudo-randomly distributed local modifications. Network covert channels utilize network packets as covert carriers. Compared to steganographic techniques that employ multimedia files like images, network covert channels leverage a vast number of packets, each offering a small hiding capacity. This characteristic provides many potential hiding units during communication. If the specific set of covert carriers used is randomly selected from this vast pool of available packets, even if an adversary intercepts all network packets between the communicating parties, it remains exceedingly difficult to accurately pinpoint the exact subset employed for covert data transmission.

\begin{figure}[hptb]
	\centering
	\includegraphics[scale=0.55]{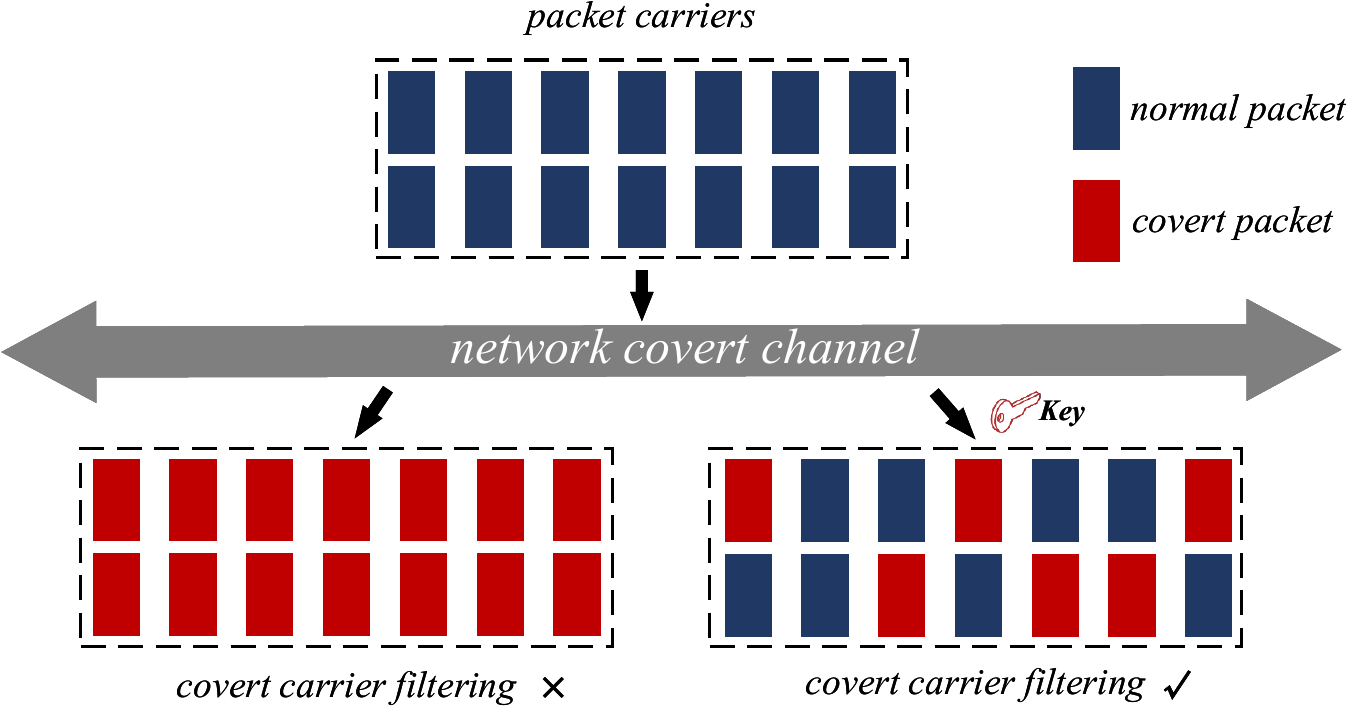}
	\caption{Covert Carrier Filtering. After covert carrier filtering, only a portion of all available carriers are used for transmitting covert data.}
	\label{covert_carrier_filtering}
\end{figure}

As shown in Fig.\ref{covert_carrier_filtering}, we introduce the covert carrier filtering step in the construction of the network covert channel. Through the filtering rules with keys, a portion of packets are filtered out from all suitable packets according to certain rules for covert data embedding. The filtering results of the filtering rules are controlled by the keys. Different keys will yield different carrier sets from the same original pool of packets. 

Under this strategy, the adversary’s challenge shifts from detecting the existence of anomalous patterns to completely and accurately enumerating all covert carriers exploited among massive volumes of packets. Although, from a theoretical standpoint, the successful identification of any single carrier implies that the existence of the covert channel has been detected, in practical scenarios achieving complete enumeration—necessary for effective disruption or precise analysis of the transmitted content—becomes exponentially more difficult. Accordingly, this strategy pursues two complementary objectives. First, by substantially increasing the operational difficulty of mounting an effective attack, it ensures that even if the covert algorithm is exposed, the adversary remains unable to accurately identify the full set of covert carriers; consequently, the channel's covertness no longer depends solely on the complexity and secrecy of the algorithm but also on the security of the key. Second, the covert carrier filtering inherently reduces channel capacity; by explicitly trading channel capacity for enhanced covertness, the overall covertness of the network covert channel is further improved. In this paper, our contributions are as follows:

\begin{enumerate}
	\item We present a formal definition and model for network covert channels, incorporating a covert carrier filtering strategy into their construction. Using information theoretical analysis, we demonstrate that this strategy enhances the network covert channel's covertness, which is shown to be dependent on the size of the key space. This analysis further informs the design requirements for effective covert carrier filtering.
	
	\item We design and implement a covert carrier filtering strategy based on SHA-256 hash. This strategy employs both an Input Key and a Filter Key, which together forms a pre-shared key between the covert sender and covert receiver (CS\&CR). The SHA-256 is selected as the filtering rule, and the covert carrier set is selected from the CS\&CR carrier set by Hash calculation and filtering of the carrier.
	
	\item We evaluate the performance of our proposed strategy in a real-world network environment for both network covert storage and timing channels. The evaluation metrics include covertness against machine learning-based traffic analysis and processing overhead. Experimental results demonstrate that with a filter key size $L$ greater than 6, the impact on the detection effect of the classifier becomes quite significant. Concurrently, the per-packet processing overhead remains below 8$\mu s$, validating the strategy's practicality for real-world deployment.
	 
\end{enumerate}

\subsection{Organization}
\label{Organization}
The remainder of this paper is organized as follows. Section \ref{Related_Work} provides a detailed review of related work on network covert channels. In Section \ref{Model}, we present a formal definition of the network covert channel model and conduct an analysis of its covertness after incorporating the covert carrier filtering. Section \ref{Strategy} introduces and implements covert carrier filtering based on the hash function. Section \ref{Experiment} evaluates the proposed covert carrier filtering through experimental analysis and testing in a real-world network environment. Section \ref{Limitations_Discussions} discusses the limitations and outlines potential directions for future research. Finally, Section \ref{Conclusion} concludes the paper.

\section{Related Work}
\label{Related_Work}

\begin{figure}[hptb]
	\centering
	\includegraphics[scale=0.4]{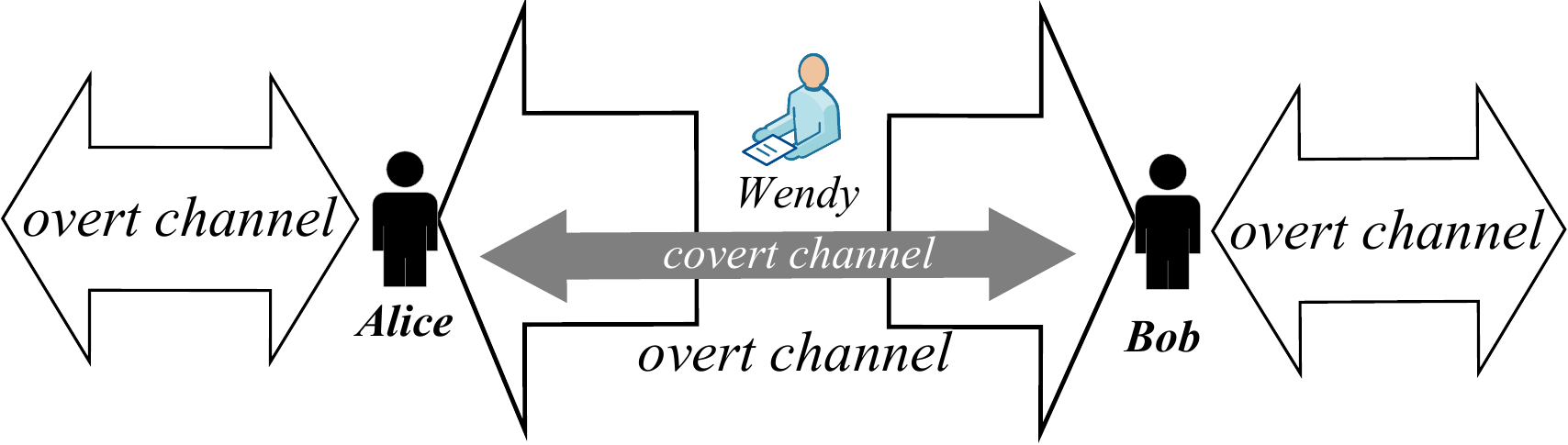}
	\caption{The Prisoner Model}
	\label{the_prisoner_model}
\end{figure}

The concept of a covert channel was first proposed by Lampson \cite{paper_3} in 1973, who defined it as a communication channel that violates security policies by transmitting information through means not originally intended for communication. The classical adversarial model is the prisoners’ problem proposed by Simmons\citep{paper_4}, as illustrated in Fig.\ref{the_prisoner_model}. In this model, Alice and Bob are two prisoners attempting to devise an escape plan. However, all their communications are monitored by the Warden. Wendy will terminate their exchanges immediately upon detecting any suspicious content. Therefore, Alice and Bob must conceal their secret messages within seemingly normal communications to evade Wendy’s surveillance. Handel et al. \citep{paper_5} extended the prisoners’ problem to network communication scenarios. Later, Girling (1987) \cite{paper_6} introduced the concept of network covert channels, which refer to communication channels in network environments where malicious parties encode and transmit information by modifying the values, characteristics, or states of shared network resources. The selection of shared network resources depends on the type of covert channel and the specific communication context; in general, network packets jointly accessible to the CS\&CR serve as the primary carriers for covert communication.

\subsection{Network Covert Channel Construction}

According to their construction mechanisms, covert channels can be classified into covert storage and timing channels \cite{paper_1_1,paper_1_2,paper_1_3}. This classification has been widely accepted by subsequent researchers and has served as the foundation for further studies. Llamas et al. \cite{paper_7} provided a detailed discussion of the construction methods for both covert storage and timing channels in network environments.

\subsubsection{Network Covert Storage Channel}

Network covert storage channels are typically constructed within packet header fields. Zander et al. \cite{paper_8} embedded secret data in the IP header’s TTL field. Because intermediate network nodes also modify TTL, the channel’s capacity and stealth depend on TTL dynamics. Hassan et al. \cite{paper_9} exploited the IPv4 timestamp field to exchange covert data and implemented the covert channel over HTTP, motivated by the high volume of HTTP traffic. Although storage channels often offer higher throughput and robustness, modifying protocol fields tends to reduce their covertness. In recent years, researchers have explored network covert channels based on packet payloads. Barradas et al. \cite{paper_10} proposed Protozoa, a WebRTC-based covert channel that replaces original WebRTC payloads with covert information via payload rewriting. Balboa \cite{paper_11} intercepts outgoing packets between an application and the operating system, compresses each packet payload to a short pointer into a pre-shared traffic model, fills the reclaimed space with covert data.

\subsubsection{Network Covert Timing Channel}

Network covert timing channels primarily encode and transmit covert data by exploiting the temporal characteristics of network packets. Tahir et al. \cite{paper_13} encoded covert data by adjusting the inter-packet delay (IPD) of normal packets sent by the CS to control their arrival times at the CR. If the CR did not receive a packet within a predefined period T, the transmitted bit was interpreted as “0”; otherwise, it was “1.” Ghassami et al. \cite{paper_14} investigated Covert Queueing Channels (CQCs), a type of timing covert channel that can operate within shared queues of so-called isolated users. Zhang et al. \cite{paper_15} proposed a covert timing channel by packet rearrangement, in which covert data is modulated into the number of Real-time Transport Control Protocol (RTCP) packets between consecutive VoLTE packets. They also employed Gray code to reduce packet loss during transmission. Network covert timing channels exhibit strong covertness but generally suffer from low robustness and transmission efficiency, as they are vulnerable to noise, network congestion, and other factors that cause packet arrival delays.

\subsection{Network Covert Channel Detection}

In the study of network covert channel detection, numerous machine learning–based approaches have been proposed. Sohn et al. \cite{paper_16} employed a Support Vector Machine (SVM) to identify network covert storage channels within the TCP/IP protocol. The selected features included the Identification field in the TCP/IP header and the Sequence field in IP packets. Both linear and polynomial kernel functions were applied to classify covert communication patterns. Bethencourt et al. \cite{paper_17} utilized neural networks trained on sequences of initial sequence numbers (ISNs) from different operating systems to detect TCP ISN–based network covert storage channels, achieving high detection accuracy in experiments. Borders et al. \cite{paper_18} focused on network covert channels embedded in the HTTP protocol and constructed a detection model using features such as request field size, inter-request interval, transmission time, and outbound bandwidth utilization. Guang et al. \cite{paper_19} proposed a joint analysis of TCP protocol field data, extracting inherent relationships among field attributes across packets. Using kernel density estimation, coefficient of variation, and autocorrelation coefficient methods, they transformed related attributes into feature vector matrices. These matrices were then classified with an SVM, achieving high detection speed while effectively reducing computational complexity.

Detection methods for network covert timing channels can generally be categorized into three major approaches: morphology based detection, regularity based detection, and entropy based detection. All these methods focus on extracting network traffic information and identifying changes or statistical anomalies in the distribution of inter-packet intervals \cite{paper_2}. In recent years, machine learning–based detection techniques for network covert timing channels have also emerged. Shresth et al. \cite{paper_20} proposed a machine learning framework for detecting network covert timing channels in network traffic. Their method employed a SVM classifier and extracted four statistical fingerprint features from the temporal characteristics of traffic: K–S statistic score, regularity score, entropy score, and modified conditional entropy score. The SVM was then trained and tested using these four types of statistical fingerprints to identify network covert timing channels. Zseby et al. \cite{paper_21} applied three density-based unsupervised learning methods that compute K-distance to detect anomalies in covert channels generated by seven different network covert timing channels. Their findings showed that, although these methods could successfully distinguish covert channels from normal traffic, they were unable to identify which specific technique had been used to create the network covert timing channel. Darwish et al.  \cite{paper_22} proposed a hierarchical statistical analysis framework combined with a deep neural network to detect covert timing channels, leveraging five statistical metrics across multiple levels of inter-arrival time flows to improve detection accuracy. Al-Eidi et al. \cite{paper_23} proposed a deep learning-based framework that leverages sequential inter-arrival time data with LSTM, 1D-CNN, and their hybrid architectures to automatically detect covert timing channels.

\subsection{Comparison}

In this paper, we propose a covert carrier filtering strategy. Rather than designing new encoding mechanisms, the strategy introduces a carrier filtering stage into the construction of network covert channels, addressing the question of which carriers should be utilized. This strategy provides a protocol-agnostic and generic framework for network covert channel construction: it does not rely on available fields at any specific protocol layer, but instead treats covert carrier filtering as an independent security primitive that can be applied to covert channel designs. Owing to this generality, the strategy is applicable to virtually all types of network covert channels and provides them with a unified, key-based security enhancement. As a concrete instantiation, we design and implement a covert carrier filtering strategy based on hash functions, and experimentally demonstrate its effectiveness.

It should be noted that covert carrier filtering or different carriers in this paper refer to the selection of a subset of packets from a continuous packet sequence, within the same covert channel, the same protocol, and the same embedding method.

As shown in Table \ref{Work_Comparsion}, we compare our approach with several related works, including those that construct network covert channels using hash functions. Liu et al. \cite{paper_1_4} proposed LaSPsteg, a covert channel scheme for LTE-A systems, which hides information by jointly exploiting the RLC layer sequence numbers and MAC layer padding bits. The CS\&CR pre-share a hash function to dynamically generate a set of sequence number values; only packets whose SNs match this set are treated as covert carriers, and their MAC padding bits are replaced with secret information. While the design philosophy of LaSPsteg shares certain similarities with our work, it essentially uses hash functions to precompute a shared set of key values for selecting packets with specific SNs as covert carriers. Although effective within its targeted domain, this scheme exhibits limited generality. Wendzel et al. \cite{paper_12} presented DYST, which leverages existing legitimate traffic: the CS monitors broadcast traffic on a local network, hashes payloads to produce fixed-length sequences, compares these sequences with the covert data to find matches, and then signals the receiver over a control channel to complete covert transmission.
	
Keller et al. \cite{paper_1_5} proposed a reversible and plausibly deniable covert channel in one-time passwords based on hash chains. This channel leverages the pseudo-random appearance of hash values as information carriers and embeds secret symbols into the hash values via XOR operations. Wang et al. \cite{paper_1_6} introduced LTCC, a covert channel over blockchain based on label tree, which embeds secret information into blockchain transaction addresses using a dynamic label-tree structure. Ma et al. \cite{paper_1_7} proposed ABC-Channel, an advanced blockchain-based covert channel aimed at supporting secure covert communication throughout the entire communication lifecycle. Partala \cite{paper_1_8} presented a provably secure blockchain covert communication scheme, in which encrypted message bits are embedded into the least significant bits of blockchain payment addresses to enable covert communication between a sender and a receiver, with security and reliability formally proven in the random oracle model.

\begin{table}[hptb]
	\centering
	\caption{Comparison of Related Work}
	\label{Work_Comparsion}
	\renewcommand{\arraystretch}{1.5}
	\begin{tabularx}{\textwidth}{>{\centering\arraybackslash}p{4cm}|>{\centering\arraybackslash}p{6cm}|>{\centering\arraybackslash}p{2cm}}
		\hline
		\textbf{Method}                     & \textbf{Role of Hash }                                                                                     & \textbf{Filtering Function} \\ \hline
		Covert Carrier Filtering Strategy   & Hash, as a protocol-agnostic filtering rule, is used to filter out covert carriers                                         & yes                                                             \\ \hline
		LaSPsteg\cite{paper_1_4}                           & Hash is used to dynamically generate a set of SN values, and the receiver identifies the covert packet by matching the SN. & yes                                                             \\ \hline
		DYST\cite{paper_12}                                & Hash is used to calculate Hash values to match covert data                                                                 & yes                                                             \\ \hline
		Covert Channel based on Hash Chains \cite{paper_1_5} & Hash is used as part of the blockchain or as computational tools                                                           & no                                                              \\ \hline
		Blockchain-based Covert Channels \cite{paper_1_6,paper_1_7,paper_1_8}    & Using the Hash value as the covert carrier, the covert data is encoded/decoded through the hash chain calculation          & no                                                              \\ \hline
	\end{tabularx}
\end{table}

\section{Network Covert Channel Model}
\label{Model}
\subsection{Model Definition}

In this section, we provide a formal description of a network covert channel. A network covert channel is defined as a five-tuple system $\mathrm{\Omega}=<C,D,\widetilde{C},\mathrm{\Phi},\mathrm{\Psi}>$, jointly specified by the CS and the CR within a network environment, where:

C (Carrier Set): The set of legitimate network packets shared between CS and CR for data transmission, denoted as $C=\left \{ c_1,c_2,c_3,\ldots \right \} $. These packets are normal resources exchanged or expected to be exchanged during legitimate communication. The CS\&CR can encode covert data by manipulating these shared resources.

D (Covert Data Set): The collection of covert data transmitted secretly from CS to CR through the network covert channel, represented as $D=\left \{d_1,d_2,d_3,\ldots \right \}$. Each covert data is embedded into covert carriers using the covert algorithm.

$\widetilde{C}$ (Covert Carrier Set): The set of modified network packets containing covert data, obtained after applying the covert algorithm, denoted as $ \widetilde{C}=\left \{{\widetilde{c}}_1,{\widetilde{c}}_2,{\widetilde{c}}_3,\ldots \right \}$.

$\mathrm{\Phi}$ and $\mathrm{\Psi}$ together denote the covert algorithms. Specifically, $\mathrm{\Phi}$ is the embedding function, used by the CS, defined as $\Phi:C\times D\rightarrow\widetilde{C}$. It encodes covert data into the carrier packets based on specific patterns or features of C(e.g., statistical characteristics, redundant fields, or behavioral traits) to produce the covert carriers $\widetilde{C}. \mathrm{\Psi}$ is the extracting function, used by the CR, defined as $\Psi:\widetilde{C}\rightarrow D$. It takes the received covert carriers as input and, by applying the inverse or decoding process corresponding to the embedding phase, reconstructs the original covert data $D$.

\begin{definition}
	A covert carrier filtering strategy refers to the process of filtering a subset of available shared resources as covert carriers to embed covert data. The selection is performed by a key-controlled filtering rule, where the key determines the outcome of the filtering process.
\end{definition}

\begin{figure}[hptb]
	\centering
	\includegraphics[scale=0.45]{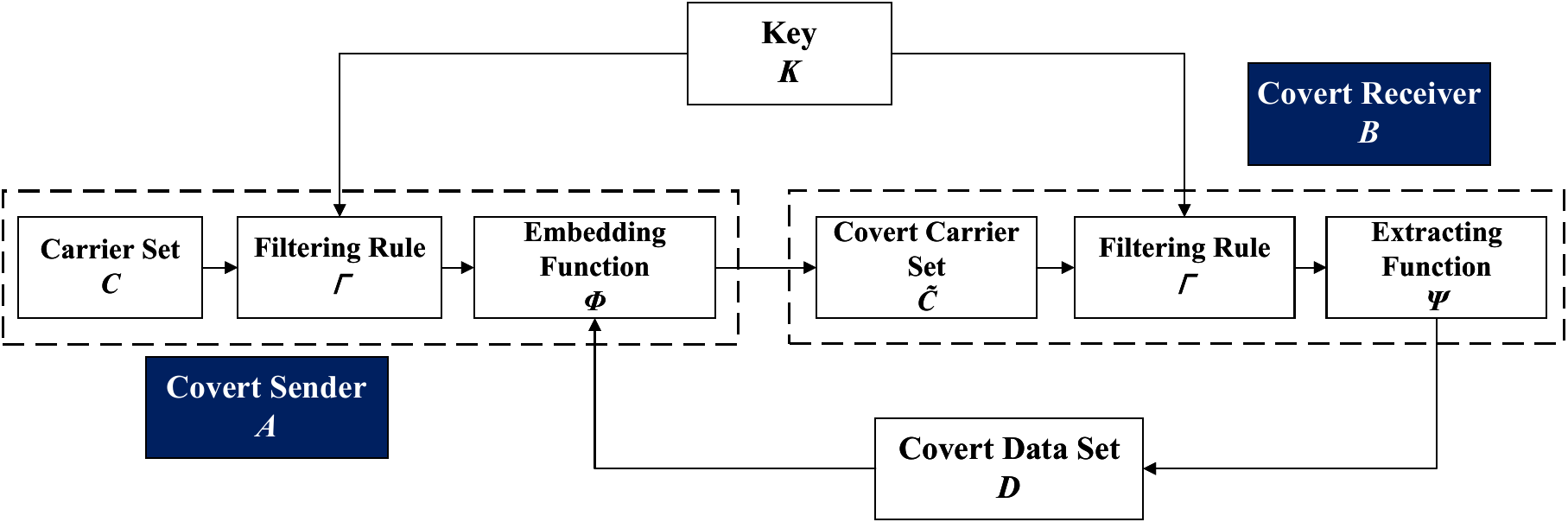}
	\caption{Network Covert Channel Model}
	\label{NCC_Model}
\end{figure}

As illustrated in Fig.\ref{NCC_Model}, a covert carrier filtering strategy is introduced into the network covert channel model. The extended model is thus defined as a seven-tuple system $\mathrm{\Omega}=<C,\widetilde{C},D,K,\Gamma,\mathrm{\Phi},\mathrm{\Psi}>$. In this model, $\Gamma$ denotes the filtering rule, which is used to filter covert carriers for embedding covert data from the carrier set $C$ and the covert carrier set $\widetilde{C}$. $K$ represents the set of possible keys used to control the filtering rule. With the filtering rule incorporated, the embedding and extracting processes are defined as follows:

at the CS,

\begin{equation}
	\mathrm{\Phi}:\Gamma(C,K)\times D\rightarrow C^\ast\times D\rightarrow\widetilde{C} ;\ C^\ast\subseteq C
\end{equation}

at the CR,

\begin{equation}
	\mathrm{\Psi}:\Gamma(\widetilde{C},K)\rightarrow{\widetilde{C}}^\ast \rightarrow D ;\ {\widetilde{C}}^\ast\subseteq\widetilde{C}
\end{equation}

The filtering rule may be public; however, the $K$ must remain secret. This design follows Kerckhoffs’s principle, which states that the security of a cryptographic system should rely solely on the secrecy of the key, while the algorithms themselves should be publicly known.

\subsection{Covert Data Security Analysis}

In a network covert channel model that incorporates a covert carrier filtering strategy, once the CS is provided with the shared resources $C$, the key $K$, and the covert data D, the corresponding $\widetilde{C}$ can be uniquely determined. From an information-theoretic perspective, this can be expressed as $H\left(\widetilde{C}|(C,K,D)\right)=0$. For the CR to correctly decode $D$, the necessary and sufficient condition is $I\left(D;\widetilde{C}|K\right)=H\left(D\right)$, which implies $H\left(D\middle|\widetilde{C},K\right)=0$. In other words, given the covert algorithm, once the CR knows both $\widetilde{C}$ and $K$, the covert data D can be uniquely determined \cite{paper_1_9}.

During the communication between the CS\&CR, the primary objective of an adversary is to obtain the covert data D by intercepting $\widetilde{C}$. In a traditional network covert channel, the design primarily focuses on the embedding function $\Phi$ and the extracting function $\Psi$. The level of covert data security mainly depends on the covertness capability and secrecy of the covert algorithm, which can be represented by the conditional entropy $H(D\mid\widetilde{C})$. In this context, when the adversary is unaware of the covert algorithm, the security of the network covert channel depends on the uncertainty of $D$ given $\widetilde{C}$. As $H(D\mid\widetilde{C})$ approaches $H(D)$, the adversary’s ability to infer $D$ from $\widetilde{C}$ diminishes, resulting in stronger security. When $H(D\mid\widetilde{C})=H(D)$, the covert data achieves perfect security, meaning that $\widetilde{C}$ provides no additional information that can help the adversary infer D. After introducing the covert carrier filtering strategy, the inclusion of the $K$ affects the information relationship such that additional information does not increase uncertainty. Therefore, $H\left(D\middle|\widetilde{C}\right)\geq H\left(D\middle|\widetilde{C},K\right)$. Since $H\left(D\middle|\widetilde{C}\right)=H\left(D\middle|\widetilde{C},K\right)+I\left(D;K|\widetilde{C}\right)$, covert data security depends on two aspects: (1) the covertness capability and secrecy of the covert algorithm, represented by $H(D\mid\widetilde{C},K)$; and (2) the randomness of the covert carrier set provided by the filtering rule, represented by the conditional mutual information $I(D;K\mid\widetilde{C})$.

When the $K$ is independent of the covert data $D$, that is, when the filtering rule is inactive, we have $I(D;K\mid\widetilde{C})=0$ and $H\left(D\middle|\widetilde{C}\right)=H\left(D\middle|\widetilde{C},K\right)$. In this case, the system degenerates into a traditional network covert channel model. Once the covert algorithm exhibits low covertness capability and is inferred or disclosed by an adversary, $H(D\mid\widetilde{C})=0$. The adversary can then uniquely determine $D$ from the observed $\widetilde{C}$, resulting in the complete exposure of the covert data.

If the $K$ provides additional information about $D$, meaning that the covert carrier set $\widetilde{C}$ is selected from $C$ according to a key-dependent filtering rule, then $I(D;K\mid\widetilde{C})>0$. When the covert algorithm is compromised, $H(D\mid\widetilde{C},K)=0$ while $H(D\mid\widetilde{C})>0$. In this case, the adversary cannot infer $D$ solely from $\widetilde{C}$, as they cannot determine which specific packets carry the covert data. Consequently, the covert data security can still be maintained through the algorithmic strength of the filtering rule.

The algorithmic strength of the filtering rule refers to its resistance to attacks—that is, the difficulty for an adversary, without knowledge of the key K, to infer or reconstruct ${\widetilde{C}}^\star=\Gamma(\widetilde{C},K)$ solely from the public information $\widetilde{C}$. A higher algorithmic strength implies a more secure network covert channel. Since $\widetilde{C}$ is given and the filtering rule $\Gamma$ is public, the variation of ${\widetilde{C}}^\star$ depends entirely on K. If $K$ is uniformly random and independent of $\widetilde{C}$, and $\Gamma$ exhibits strong pseudo randomness, then:

\begin{equation}
	I\left(D;K|\widetilde{C}\right)\approx H(K)
\end{equation}

Let $H(K)$ denote the entropy of the $K$, representing the size and randomness of the key space. A larger value of $I(D;K\mid\widetilde{C})$ indicates that even if the adversary observes $\widetilde{C}$, they cannot determine the set ${\widetilde{C}}^\star$. Ideally, $I(D;K\mid\widetilde{C})$ should approach $H(K)$, thereby maximizing uncertainty. In this scenario, the adversary would need to perform an exhaustive search over the entire key space to compromise $\widetilde{C}$.

Network covert channels utilize network packets as covert carriers. Compared to information hiding techniques in multimedia files such as images, network covert channels handle larger volumes of carrier data while hiding smaller units of information. This characteristic imposes the following design requirements on the filtering rule $\Gamma$:

\begin{enumerate}
	\item \textbf{Key Dependence.} $\Gamma$ must ensure that the selection of $\widetilde{C}$ strongly depends on the $K$. If $\Gamma$ is insensitive to K, it will reduce the covert data security.
	
	\item \textbf{Output Uniformity.} The rule should ensure that ${\widetilde{C}}^\star$ is uniformly distributed within $\widetilde{C}$. preventing concentration in specific positions; otherwise, the adversary could narrow the monitoring scope.
	
	\item \textbf{Computational Efficiency.} The filtering rule should possess high computational efficiency to minimize the time required for covert carrier selection and avoid interfering with normal network operations.
	
\end{enumerate}

\section{Hash-Based Covert Carrier Filtering Strategy}
\label{Strategy}

This section first analyzes the adversarial threats faced by network covert channels, then elaborates on the design and implementation of a hash–based covert carrier filtering strategy, and finally introduces the method for pre-shared key negotiation between the CS\&CR.

\subsection{Threat Model}

In this paper, we assume that the CS\&CR are legitimate communication users, and they use the network packets directly interacting with the upper application to construct an end-to-end network covert channel to transmit covert data, for example, monitoring devices and monitoring clients in the same local network, application clients and remote servers. The CS\&CR share a predefined secret key and employ a covert carrier filtering strategy as a means of enhancing covertness.

The adversary is modeled as a passive warden, who can only monitor the communication channel but cannot modify any transmitted information. Specifically, the adversary can intercept and store intermediate traffic between the CS\&CR, as well as analyzing metadata features such as packet size distribution, timing intervals, and traffic rates. Furthermore, the adversary may employ traffic analysis techniques, including machine learning, to detect anomalous flows.

This paper discusses the application of the covert carrier filtering strategy in two typical scenarios.

\textbf{Scenario 1} — Network storage covert channels: the CS\&CR embed covert data by modifying redundant or optional fields within network protocol packets. Typical candidate fields include the IP identification field, Differentiated Services Field (DS), TCP sequence number, and HTTP header fields. Conventional approaches are vulnerable for two reasons. First, modifying a fixed set of carrier fields alters the global traffic statistics, making the covert traffic detectable by machine learning–based classifiers. Second, researchers have largely enumerated the exploitable fields, so network storage covert channels are increasingly susceptible to rule-based detection targeted at these fields. By introducing covert carrier filtering strategy, modifications are dispersed across packets rather than concentrated locally: the covert traffic then contains a mixture of modified and unmodified packets. The large proportion of unmodified legitimate packets “dilutes” statistical anomalies so that the overall distribution increasingly approximates normal traffic, thereby evading distribution-based classifiers. Moreover, even if the adversary knows the covert algorithm, without the secret key they cannot reliably identify which packets contain covert data among large volumes of network traffic. Thus, covertness shifts from algorithm secrecy to key strength in accordance with Kerckhoffs’s principle: the adversary’s task changes from pattern recognition to key search, incurring exponential computational cost. 

\textbf{Scenario 2} — Network timing covert channels: the CS\&CR construct a network timing covert channel by encoding information in packet-timing characteristics (e.g., IPDs, packet order, or presence/absence) rather than by modifying packet contents. The covert channel’s covertness depends on being masked by the network’s inherent jitter. Typical carriers include ICMP request–response intervals, TCP packet arrival times, and the ordering of request and response packets. Conventional network timing covert channels that employ fixed modulation patterns introduce anomalous statistical regularities and are therefore vulnerable to time-series analysis tools. By applying a covert carrier filtering strategy, modulation events are made aperiodic and pseudo-randomly distributed along the time axis, disrupting any fixed modulation periodicity and preventing the time series of covert traffic from exhibiting regularities. It should be clear that the network timing covert channel with IPD as the modulation target involves two packets ordered, $P1$ and $P2$. The filtering object in this strategy refers to $P1$, which is selected by the filtering rule. We take the IPD between $P1$ and $P2$ as the modulation object to encode the covert data. The strategy can be coupled to current network conditions to dynamically adjust modulation amplitude and filtering thresholds, thereby avoiding outliers that would result from inserting high-amplitude modulation during low-jitter periods and improving resistance to anomaly-detection methods.

The proposed covert carrier filtering strategy, as an enhancement mechanism for network covert channels, is fundamentally distinct from existing network covert channel construction patterns \cite{paper_1}, particularly Distributed Covert Channels \cite{paper_1_10} and their sub-categories (e.g., Flow-based scattering). Such approaches aim to increase analytical difficulty by introducing greater pattern complexity across different contexts; however, their essence still lies in variations of data encoding patterns. In contrast, the filtering strategy does not directly modify the data encoding pattern itself. Instead, it introduces an upstream, cryptography-driven filtering layer that governs whether covert data encoding is applied to a given packet. As an encoding-independent and more fundamental carrier management mechanism, the strategy is compatible with most construction patterns—such as PT1, PS1, PS3, PS10, PS20, and PS30 \cite{paper_1_10}—by providing an additional layer of security protection. Moreover, different construction patterns induce distinct types of statistical anomalies. By deliberately foregoing modification opportunities for the majority of carriers, the strategy allows a large volume of unmodified packets to dilute the microscopic anomalies introduced by localized modifications. The objective is to make the macroscopic traffic distribution asymptotically converge to the original background traffic, rather than to generate a new, more complex distribution.

An overview of the notation system used in this paper is presented in Table \ref{Notations_Table}.

\begin{table}[hptb]
	\centering
	\caption{Table of Used Notations}
	\label{Notations_Table}
	\renewcommand{\arraystretch}{1.2}
	\begin{tabular}{c|p{10.5cm}}
		\hline
		\textit{\textbf{Symbol}} &  \textit{\textbf{Definition}}                                                                                 \\ \hline
		$C$                 & Carrier set, the network packet resources shared by both the CS\&CR      \\ \hline
		$c_i$               & The i-th packet in the carrier set						  \\ \hline
		$\widetilde{C} $    & Carrier set after embedding covert data through covert algorithms  \\ \hline
		${\widetilde{c}}_i$ & The i-th packet in the carrier covert set      \\ \hline
		$D$                 & Covert data set, secret information to be transmitted       \\ \hline
		$d_i$  				& The i-th bit covert data in the covert data set         \\ \hline
		$K$ 				& The secret key is shared between the CS\&CR  \\ \hline
		$Input\ Key$ 		& The key of the first part of K, which is used for the Hash function input \\ \hline
		$Filter\ Key$		& The key of the second part of K, which is used for Hash value filtering   \\ \hline
		$L$ 				& Filter Key Size  \\ \hline
		$Hash(\cdot)$			& Cryptographic Hash functions use $\cdot$ as input \\ \hline
		$h_i$ 				& Hash value of the i-th packet \\ \hline
		$r$ 				& Proportion of covert carriers in normal carriers \\ \hline
		$e$ 				& Coding efficiency of covert algorithms \\ \hline
		$U_{cc}$ 			& Unit Covert Capacity \\ \hline
		$t_{average}$ 		& Average Per-Packet Processing Time \\  \hline
		
	\end{tabular}
\end{table}

\subsection{Design and Implementation}

This section describes the design and implementation of a hash-based covert carrier filtering strategy. Hash functions are a fundamental technique in cryptography. They efficiently map inputs of arbitrary length to fixed-length, one-way outputs and are designed to be collision-resistant and highly sensitive to input changes. The resulting value—commonly called a hash value, message digest, or fingerprint—is well suited for constructing filtering rules. Based on these properties, using hash-based rules to filter covert carriers offers three main advantages:

\begin{enumerate}
	\item By using a pre-shared key together with designated packet fields as the hash input, the function’s input sensitivity ensures that even small differences in input produce very different hash outputs.
	
	\item Hash-based filtering yields an effective uniform and pseudo-random mapping from packets to covert carriers; without the key, an adversary cannot predict which packets will be chosen, making it difficult to identify the covert carrier set.
	
	\item Hash algorithms are computationally efficient and impose minimal processing overhead, allowing rapid covert carrier filtering in high-volume traffic scenarios.
\end{enumerate}

\subsubsection{Implementation Steps}

\begin{figure}[hptb]
	\centering
	\includegraphics[scale=0.45]{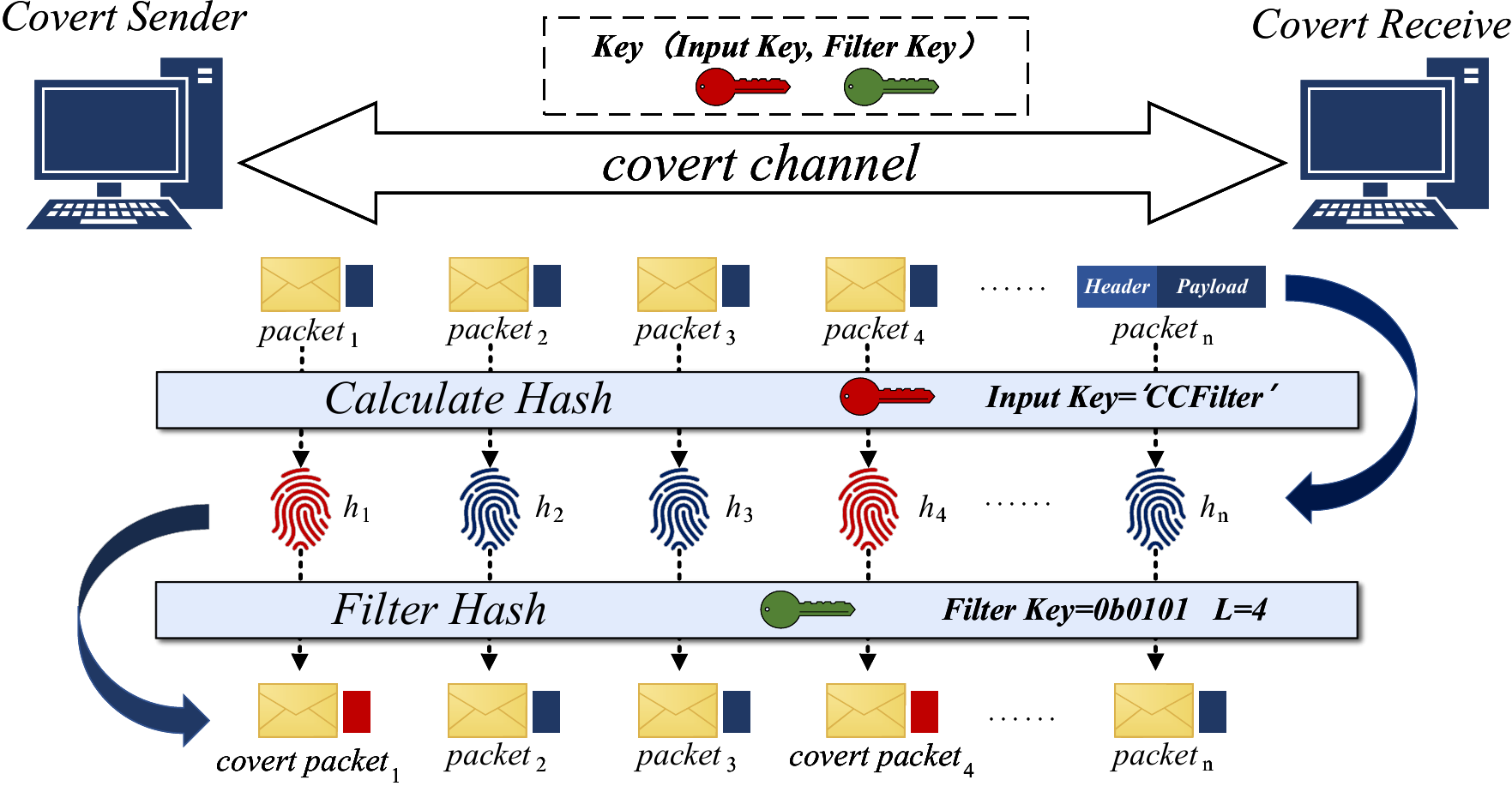}
	\caption{Hash–Based Covert Carrier Filtering Process.}
	\label{filtering_process}
\end{figure}

The purpose of the covert carrier filtering strategy is to enable the CS\&CR to use the same filtering rule to extract an identical covert carrier set $\widetilde{C}=\left \{{\widetilde{c}}_1,{\widetilde{c}}_2,\ldots,{\widetilde{c}}_m \right \}$, from the set of normal traffic $C=\left \{c_1,c_2,\ldots,c_n \right \}(with\ m<n)$, thereby ensuring carrier synchronization between the CS\&CR. The secrecy of the pre-shared key guarantees that the filtering outcome remains confidential and increases the difficulty for an adversary to discover $D$. As illustrated in Fig.\ref{filtering_process}, the hash–based covert carrier filtering strategy executes the following steps to obtain the covert carrier set:

\begin{enumerate}
	\item \textbf{Pre-Shared Key Setup:} The CS\&CR pre-share a $K=\left \{ Input\ Key ,\right.$ \\ $\left. Filter\ Key \right \} $, composed of an input key and a filter key. The input key participates in hash computation as one of the inputs to the hash function, while the filter key is used to select packets whose hash values match a specified criterion, designating them as covert carriers.
	
	\item \textbf{Hash Computation:} The CS intercepts network packets and computes the hash value of each packet payload as $h_i=\mathrm{Hash}(\mathrm{Input\ Key},c_i)$ using the pre-shared hash function and input key. To ensure unambiguous synchronization between the CS\&CR, the packet features used for hash computation must remain unchanged during transmission. Therefore, the packet payload is chosen as the input, as it is generally not modified by intermediate network devices. Source and destination IP fields are avoided because network address translation may alter them, which would lead to inconsistent hash outputs.
	
	\item \textbf{Covert Carrier Filtering:} The CS applies the filter key to select packets based on the hash values. A packet $c_i$ is selected as a covert carrier if $(h_i \& mask)==\mathrm{Filter\ Key}$, where $mask=2^L-1$. The selection uses a bitmask operation for high computational efficiency, minimizing per-packet processing time. Essentially, filtering compares the least significant $L$ bits of the hash value with the filter key. From a cryptographic perspective, since the hash output is uniformly distributed, its least significant $L$ bits can also be regarded as uniformly random over $[0,2^L-1]$, ensuring that the filtering is random.
	
\end{enumerate}

It should be noted that cryptographic hash functions are inherently unkeyed cryptographic primitives. To achieve key-dependent filtering decisions, we concatenate the Input Key with the packet payload as the input to the hash function. This makes the filtering outcome predictable only to communicating parties possessing the same key, while preserving the pseudorandom property of the hash function.

Furthermore, for network covert channels that use payload fields as embedding carriers (such as certain types surveyed in \cite{paper_1_10}), our strategy cannot be directly applied. This is because modifying the payload will cause the hash values calculated by the CS\&CR to be inconsistent, thereby disrupting the synchronization of the filtering process.

\subsubsection{Parameter Settings}

In the covert carrier filtering process, the input key serves as one of the inputs to the hash function along with the packet payload, ensuring that an adversary cannot compute the same hash value without knowledge of the key. The filter key is used to select packets based on their hash values, specifically by checking whether $h_i \& (2^L-1)=Filter\ Key$ $\left ( L\ge 0 \right ) $. The filter key size $L$ is variable: a longer filter key requires more bits to match, reducing the probability of a match and decreasing the proportion of covert carriers.
Thus, the filter key simultaneously controls the proportion of covert carriers that are filtered. The relationship between the filter key size and the filtering ratio $r$ is given by:

\begin{equation}
	r=\frac{1}{2^L}
\end{equation}

This indicates that the filtering process can be modeled as a Bernoulli trial with parameter $r$, where the occurrence of covert carriers among candidate packets is independent and random. When $L$ is 0, $r$ equals 1. All candidate data packets are used as covert carriers, which means no screening strategy is adopted. 

\subsubsection{Workflow}

In our design, the SHA-256 hash function is used as the core component of the filtering function. Its fixed 256-bit output, strong collision resistance, and one-way property ensure the randomness and security of the filtering process. The covert carrier filtering strategy serves as a step in constructing a network covert channel, with the ultimate goal of enabling covert data transmission. We assume that the CS\&CR establish a network storage covert channel, embedding covert data within protocol field redundancies. The covert data transmission proceeds as follows:

1. \textbf{Parameter Coordination}. Before constructing the network covert channel, the CS\&CR must coordinate the following parameters: the SHA-256 hash function, the pre-shared key $K=\left \{Input\ Key,Filter\ Key \right \}$, and the covert algorithm.

2. \textbf{CS's operations}. The CS monitors the network interface and captures candidate packets $c_i$. For each packet $c_i$, the CS extracts the payload and computes its hash value using the SHA-256 hash function. Packets satisfying the filtering rule are selected as covert carriers ${\widetilde{c}}_i$. The CS then embeds covert data into the covert carriers using the embedding algorithm and transmits them to the CR over the overt channel. The implementation is as shown in Algorithm \ref{Algorithm_1}.

\begin{algorithm}[!h]
	\caption{CS Implementation}
	\label{Algorithm_1}
	\renewcommand{\algorithmicrequire}{\textbf{Input:}}
	\renewcommand{\algorithmicensure}{\textbf{Output:}}
	\begin{algorithmic}[1]
		\REQUIRE $C=\left \{ c_1,c_2,\ldots \right \} $, $ K=\left \{Input\ Key, Filter\ Key\right \}$, $ D=\left \{d_1,d_2,\ldots \right \}$
		\ENSURE $\widetilde{C}=\left \{{{\widetilde{c}}_1,{\widetilde{c}}_2,{\widetilde{c}}_3,\ldots}\right \}$    
		
		\STATE ${mask=2}^L-1$
		\FOR{$captured\ packet\ c_i\ in\ traffic$}
		\STATE $feature\gets\mathrm{Extractpayload}\left(c_i\right)$
		\STATE $h_i\gets Hash(Input\ Key\ \ ||\ \ feature)$
		\IF {${(h}_i\ \&\ mask)==Filter\ Key$}
		\STATE ${\widetilde{c}}_i \gets EmbedData\left({c_i,\ d}_i\right)$
		\STATE $Send\ {\widetilde{c}}_i$
		\ELSE  
		\STATE $Send\ c_i $
		\ENDIF 
		\ENDFOR
	\end{algorithmic}
\end{algorithm}

3. \textbf{CR's operations}. The CR performs the same operations: for each candidate packet $c_i$ received from the CS, the CR computes the hash value and applies the filtering rule to identify covert carriers and then extracts the covert data from the covert carriers according to the extracting algorithm. The implementation is as shown in Algorithm \ref{Algorithm_2}.

\begin{algorithm}[!h]
	\caption{CR Implementation}
	\label{Algorithm_2}
	\renewcommand{\algorithmicrequire}{\textbf{Input:}}
	\renewcommand{\algorithmicensure}{\textbf{Output:}}
	\begin{algorithmic}[1]
		\REQUIRE $C=\left \{ c_1,c_2,\ldots \right \} $, $ K=\left \{Input\ Key, Filter\ Key\right \}$
		\ENSURE $ D=\left \{d_1,d_2,\ldots \right \}$   
		
		\STATE ${mask=2}^L-1$
		\FOR{$captured\ packet\ c_i\ in\ traffic$}
		\STATE $feature\gets\mathrm{Extractpayload}\left(c_i\right)$
		\STATE $h_i\gets Hash(Input\ Key\ \ ||\ \ feature)$
		\IF {${(h}_i\ \&\ mask)==Filter\ Key$}
		\STATE $d_i\gets ExtractData\left(c_i\right) $
		\STATE $D\ append(d_i)$
		\ENDIF 
		\ENDFOR
	\end{algorithmic}
\end{algorithm}

Through the above process, the CS\&CR can synchronously identify the same sparse, pseudo-random set of covert carriers from a continuous network traffic, without requiring any in-band synchronization signaling. The advantage of this design lies in shifting the channel’s covertness from being dependent on the covert algorithm to being dependent on the secrecy of the key, thereby significantly enhancing covertness against traffic analysis.

\subsubsection{Pre-Shared Key Agreement}

In the design presented here, we assume that the CS\&CR have a pre-shared key $K={Input\ Key,\ Filter\ Key}$ to focus on evaluating the effectiveness of the covert carrier filtering strategy. In a practical, deployable covert communication system, however, secure key establishment and dynamic key management are essential to long-term security. This section describes feasible methods for key agreement between the CS\&CR and strategies to improve the covertness strength of the network covert channel.

\paragraph{Out-of-band agreement based on key-exchange methods}

The most direct and reliable method for establishing a pre-shared key is to use an auxiliary channel that is not monitored by the adversary.

\begin{enumerate}
	
	\item Physical exchange. Prior to commencing covert communication, the parties meet in person (for example, to exchange a USB device containing the key) or use a trusted courier to transfer an initial master key.
		
	\item Use of an existing secure channel. If the parties already share a trusted communication channel, they may transport the key over that channel. This approach anchors the covert channel’s key security in a separate, proven security protocol.
	
\end{enumerate}

\paragraph{Online agreement over the network covert channel}

When out-of-band agreement is infeasible, the network covert channel itself can be used to bootstrap a shared key. To preserve secrecy, such online agreements should employ public-key cryptographic protocols.

\begin{enumerate}

\item PKI-assisted key transport. The CS encrypts a pre-shared key with the CR’s public key and embeds the ciphertext into network storage or timing covert carriers. The CR extracts the ciphertext and decrypts it with its private key to obtain the pre-shared key.

\item Diffie–Hellman(DH) key exchange. The CS\&CR perform a standard DH exchange and hide the transmitted values (e.g., $g^a\ mod\ p, g^b\ mod\ p$) within the covert channel. Each party then derives the shared secret $s=g^{ab}\ mod\ p$ locally.

\end{enumerate}

\paragraph{Strategies to strengthen covertness}

To ensure resilience against strong adversaries, we recommend the following system-level hardening measures:

\begin{enumerate}
	
	\item Maximize key entropy and search space. The covertness of the channel depends on key unpredictability. The Input Key should be long enough to resist brute-force attacks; we recommend a minimum length of 128 bits. The effective space of the Filter Key is determined by $L$, and its values should be sampled uniformly at random.
	
	\item Choose strong cryptographic primitives. The filtering function’s security depends on the hash function’s strength. Prefer hash functions with strong collision resistance and good pseudo-random properties. In addition to SHA-256 (used in this paper), consider SHA-3 or other modern hash families where resources permit; such choices improve future resistance to cryptanalysis and help preserve uniformity of filtering outputs.
	
	\item Enforce dynamic key and policy updates. Static keys and filtering rules become more vulnerable to statistical analysis over long exposure periods. The CS\&CR should regularly refresh keys. In addition, filtering parameters (for example, the filter-key length $L$) may be varied periodically to adjust $r$, causing the channel’s statistical profile to drift over time.
	
\end{enumerate}

\section{Experiment and Analysis}
\label{Experiment}

This section evaluates the covert carrier filtering strategy under two different scenarios: network storage and timing covert channels. After describing the implementation of the filtering strategy in each scenario, along with the experimental setup and evaluation methodology, we conduct an analysis of the channel’s covertness and capacity to validate the effectiveness of the proposed strategy.

\subsection{Objectives and Methodology}

The goals of our experiment have two aspects: 1) To evaluate the impact of the covert carrier filtering strategy on the covertness of the network covert channel. 2) To investigate how the filter key size $L$ affects both the channel’s covertness and its capacity, and to identify the corresponding trade-off.

\subsubsection{Experimental Scenarios}

To comprehensively evaluate the applicability and effectiveness of the hash-based covert carrier filtering strategy, we constructed experimental environments and conducted performance tests for the two primary types of covert channels: network storage and timing covert channels.

\paragraph{Network Storage Covert Channel}

In the implementation of the network storage covert channel, we selected the DS field and ID field in the IP header as the embedding fields. This field is often ignored in normal network communication or used for simple QoS management, the field value distribution is relatively consistent, and modifying it will change its statistical characteristics. At the same time, we select the IP ID field, which should be highly random in normal communication. The traditional covert channel will destroy this randomness and make its distribution tend to be fixed, so it is easy to be detected. The specific implementation is summarized in Table \ref{NSCC}.

\begin{table}[hptb]
	\centering
	\caption{Network Storage Covert Channel Implementation}
	\label{NSCC}
	\renewcommand{\arraystretch}{1.2}
	\begin{tabularx}{\textwidth}{>{\centering\arraybackslash}p{3cm}|>{\arraybackslash}p{10cm}}
		\hline
		\textbf{Embedded field}      & The DS field and ID field in the IPv4 header.                                                                                                                                                    \\ \hline
		\textbf{Embedding algorithm} & We divide the covert data to be transmitted into consecutive 8 bit groups and randomly replace each group with the DS field or the lower 8 bits of ID field. \\ \hline
		\textbf{Embedded capacity}   & Each IP data packet selected as the carrier can hide 8 bits of covert data.   \\ \hline
	\end{tabularx}
\end{table}

This scenario aims to verify whether, after introducing the covert carrier filtering strategy, the sparse and random distribution of modifications can effectively dilute the statistical anomalies in the DS field and ID field values thereby evading detection.

\paragraph{Network Timing Covert Channel}
In the network timing covert channel implementation, covert data is encoded by precisely controlling packet transmission intervals. The network’s inherent transmission delay jitter provides natural covertness for this timing modulation. The specific implementation is summarized in Table \ref{NTCC}.

\begin{table}[hptb]
	\centering
	\caption{Network Timing Covert Channel Implementation}
	\label{NTCC}
	\renewcommand{\arraystretch}{1.2}
	\begin{tabularx}{\textwidth}{>{\centering\arraybackslash}p{3cm}|>{\arraybackslash}p{10cm}}
		\hline
		\textbf{Modulation carrier}      & The IPDs between consecutive  packets.                                                                                                                                                       \\ \hline
		\textbf{Embedding algorithm} & Differential interval coding is adopted. We defined a base time interval, $base\_interval=5ms$. When encoding, if the bit of the covert data to be sent is '0', the IPD of the next packet needs to be less than $base\_interval$. If the bit is '1', the IPD of the next packet needs to be greater than $base\_interval$. The CR can decode the covert bit by measuring the arrival time interval and comparing it with $base\_interval$. \\ \hline
		\textbf{Embedded capacity}   & Each modulated IPD can carry 1 bit of covert data.     \\ \hline
	\end{tabularx}
\end{table}

This scenario aims to verify whether, after introducing the covert carrier filtering strategy, the non-continuous, pseudo-random timing modulation can break the high autocorrelation and regularity inherent in fixed-period modulation, causing the statistical characteristics of the covert traffic’s time series to closely approximate those of normal network traffic affected only by natural jitter.

\subsubsection{Feature Extraction}

To effectively evaluate the covertness of the network covert channel, we adopt a machine learning–based detection method as the evaluation approach. This method extracts statistical features from network traffic that can indicate the presence of a network covert channel and constructs a classifier to automatically distinguish between normal traffic and covert traffic. The following sections describe the feature-extraction schemes for network storage and timing covert channels, respectively.

\paragraph{Network Storage Covert Channel}

For the network storage covert channel, the core features lie in the perturbations of protocol field value distributions. Guangxin Fu et al.\cite{paper_19} selected the following four key statistical features to construct a feature vector that captures the distributional changes caused by covert data embedding:

\subparagraph{Kernel Density Estimation ($KDE$)} 
Used to non-parametrically estimate the probability density function of protocol field values, allowing sensitive detection of subtle changes in distribution shape.

\begin{equation}
	\hat{f}\left(P\right)=\frac{1}{nh}\sum_{i=1}^{n}{K(\frac{P_i-P}{h})}
\end{equation}

\subparagraph{Coefficient of Variation ($C_v$)} 
Reflects the degree of data dispersion. It captures not only the statistical regularities within individual packet fields but also the differences among fields across packets.

\begin{equation}
	C_v=\frac{\sigma}{\mu}
\end{equation}

\subparagraph{Entropy ($H$)} 
Measures the uncertainty in the process. Higher entropy indicates that the packet header fields contain more information.

\begin{equation}
	H\left(P_1,P_2,\ldots P_n\right)=-\sum_{i=1}^{n}{p_i\log{p_i}},i=1,2,\ldots,n
\end{equation}

\subparagraph{Autocorrelation Coefficient ($R\left( \tau \right)$)} 

Describes the similarity of data across different time points. A higher autocorrelation coefficient indicates greater similarity among protocol field values.

\begin{equation}
	R\left( \tau \right)=\frac{E\left [ \left ( P_i-\mu \right ) \left ( P_{i+}-\eta  \right )  \right ] }{\sigma ^2} ,i=1,2,…,n
\end{equation}

Finally, the feature vector for a single protocol field in a network storage network covert channel is constructed as:$V=(\hat{f}\left(P\right),C_v,H,R\left(\tau\right))$. This vector provides a comprehensive description of the traffic characteristics across four dimensions: distribution shape, dispersion, randomness, and temporal correlation.

\paragraph{Network Timing Covert Channel}

For the network timing covert channel, the features primarily manifest in the dynamic characteristics of packet timing. Shrestha et al. \cite{paper_20} extracted the following four key features to capture anomalies in timing patterns:

\subparagraph{Kolmogorov–Smirnov Test ($D$)} 
Measures the maximum difference between the cumulative distribution function (CDF) of the covert traffic and that of normal traffic, determining whether the two originate from the same distribution.

\begin{equation}
	D={sup}_x|F_1\left(x\right)-F_2\left(x\right)|
\end{equation}

\subparagraph{Regularity Score ($Reg$)} 
Quantifies the degree of variation within the traffic flow.

\begin{equation}
	Reg=StdDev\left(\frac{\sigma_i-\sigma_j}{\sigma_j}\right),i<j,\forall i,j
\end{equation}

\subparagraph{Entropy Score ($H$)} 
Similar to the network storage covert channel, this evaluates the randomness of the inter-packet interval sequence.

\begin{equation}
	H\left(X_1,X_2,\ldots X_n\right)=-\sum_{i=1}^{n}{p_i\log{p_i}},i=1,2,\ldots,n
\end{equation}

\subparagraph{Corrected Conditional Entropy Score ($CCE$)} 
Measures the linear dependency and complexity of the inter-packet interval sequence, making it highly sensitive to timing-modulated covert channels.

\begin{equation}
	CCE=H\left(X_i|X_{i-1},\ldots X_n\right)+p\left(X_i\right)\bullet H(X)
\end{equation}

Finally, the feature vector for the timing-based channel is constructed as:$V=(D,Reg,H,CCE)$. This vector characterizes the traffic timing features across four dimensions: distribution consistency, regularity, randomness, and temporal complexity.

\paragraph{Classifier Design and Training}

We use three machine learning detection methods—SVM, Random Forest, and XGBoost—as covert channel detectors, similar methods have been successful in previous work \cite{paper_10,paper_11,paper_1_11}. By comparing detection rates under different network covert channel configurations—such as enabling or disabling filtering strategies and varying filter key sizes—the covertness gain can be quantitatively evaluated.

\subsubsection{Evaluation Metrics}

We employ the following metrics to evaluate the effectiveness of the covert carrier filtering strategy in enhancing the covertness of network covert channels.

\paragraph{Covertness Metrics}
Covertness is the primary indicator for assessing the survivability of network covert channels. Consistent with most existing evaluation methods for network covert channel covertness, we adopt the True Positive Rate (TPR), False Positive Rate (FPR), and the Area Under the Receiver Operating Characteristic Curve (AUC) to quantitatively measure the resistance of a network covert channel to detection. The TPR measures the proportion of covert traffic correctly identified by the detection model among all actual covert samples, whereas the FPR measures the proportion of normal traffic incorrectly classified as covert. To comprehensively assess detection performance, we plot the ROC curve and use the area under this curve(AUC) as the core evaluation metric. The ROC curve illustrates the trade-off between TPR and FPR under different classification thresholds. The AUC value quantitatively represents the probability that the classifier ranks a randomly chosen positive sample higher than a negative one, ranging from 0 to 1. When the AUC value equals 0.5, the classifier performs no better than random guessing and therefore has no predictive power. Consequently, an ideal network covert channel should aim to make the classifier’s AUC approach 0.5, indicating that its covert traffic is nearly indistinguishable from normal traffic in the feature space.

\begin{equation}
	TPR=\frac{TP}{TP+FN}, 
	FPR=\frac{FP}{FP+TN}
\end{equation}

Here, $TP$ denotes the number of covert traffic samples correctly classified, $FN$ the number of covert samples misclassified as normal, FP the number of normal samples misclassified as covert, and TN the number of normal samples correctly identified.

\paragraph{Communication Performance Metrics}

\subparagraph{Unit Covert Capacity($U_{cc}$)}
To quantify the capacity cost introduced by the covert carrier filtering strategy, we define the unit covert capacity $U_{cc}$, which measures the number of covert packets required to transmit one bit of covert data:

\begin{equation}
U_{cc}=\frac{Number\ of\ covert\ carriers\ required}{Number\ of\ covert\ data\ bits\ transmitted}\propto\frac{r}{e}
\end{equation}

For example, in a network timing covert channel employing inter-packet timing modulation, embedding one bit of data requires at least $U_{cc}=2$. In contrast, using the last two bits of the DS field requires at least $U_{cc}=0.5$. Because the introduction of covert carrier filtering prevents the capacity of a single packet from representing overall channel capacity, this metric’s core value lies in evaluating the effective capacity after filtering. Specifically, $U_{cc}$ is proportional to the filtering ratio $r$ and inversely proportional to the encoding efficiency $e$. The filtering strategy improves covertness by sacrificing carrier availability (reducing r). Hence, the overall channel capacity results from the joint filtering ratio $r$ and encoding efficiency $e$. A smaller $U_{cc}$ indicates a larger effective channel capacity, and vice versa.

\subparagraph{Average Per-Packet Processing Time($t_{average}$)}

This metric measures the average additional computation time introduced by the covert carrier filtering for each packet, including: 1) the delay caused by computing the packet hash value and comparing it with the filtering key; 2) the time required to embed or extract covert data from the covert carriers. 

\begin{equation}
	t_{average}=\frac{Total \ filtering \ time+Total\  embedding/extraction\  time}{Number \ of \ packets}
\end{equation}

$t_{average}$ directly determines whether the proposed strategy can be applied in latency-sensitive network environments without affecting normal functionality, making it a key metric for evaluating practicality and scalability.

\subsection{Environment and Dataset}

\subsubsection{Configuration}

The experiments were conducted on a hardware platform equipped with a 12th-generation Intel® Core™ i5-12400 processor (2.50 GHz) and 64 GB of RAM. Experiments were executed on Windows 10. The experimental code was developed in Python 3.9 and depends on key software libraries such as Scapy, Scikit-learn, and dpkt to build network covert channel and detection system. The parameters of the three classifiers were all adjusted to the optimal state based on preliminary experiments.

\subsubsection{Dataset Collection}

All datasets used in this paper were derived from real network traffic captured in operational environments to realistically simulate the deployment and detection scenarios of network covert channels. To account for the differing characteristics of network storage and timing covert channels, we collected two representative background traffic datasets to serve as covert carriers.

\paragraph{Network Storage Covert Channel Dataset}

We captured the continuous web browsing traffic between the local host and the remote server in an active office network environment. This dataset is predominantly composed of IP packets with TCP payload and includes HTTP/HTTPS requests and responses. The traffic exhibits complete protocol fields and dynamically varying traffic patterns, making it well suited as a covert carrier for network storage covert channels. A total of 1000,000 IP packets with TCP payload were captured. The dataset was evenly partitioned into two subsets:

\begin{itemize}
	\item Normal traffic set: Used as the baseline for training the classifier’s normal class.
	
	\item Covert traffic candidate set: Used to construct the network storage covert channel. On this subset, we applied the proposed hash-based covert carrier filtering strategy and covert algorithm to embedding covert data in the DS field, thereby generating covert traffic for evaluation.
	
\end{itemize}

\paragraph{Network Timing Covert Channel Dataset}

To emulate scenarios that tolerate macro-level timing stability and are insensitive to minor jitter, we captured real-time streaming media traffic between video surveillance devices and video surveillance client in the local area network. This dataset is primarily composed of IP packets with UDP payload. Packet inter-arrival times are relatively stable at the macro scale and packet payloads are comparatively large; however, micro-scale jitter due to network conditions is present, providing an ideal camouflage environment for network timing channels. A total of 1,600,000 IP packets with UDP payload were captured. These were equally partitioned into:

\begin{itemize}
	\item Normal traffic set: Used to establish baseline temporal features for normal traffic.
	
	\item Covert traffic candidate set: Used to construct network timing covert channels. Based on this traffic, covert timing patterns were created by controlling UDP packet transmission intervals to encode covert data, enabling evaluation of the strategy’s effectiveness in the temporal domain.
	
\end{itemize}

\subsection{Experimental Results and Analysis}

\subsubsection{Robustness experiment}

In real-world network environments, packet loss, packet reordering, and delay jitter are ubiquitous phenomena. From a mechanistic perspective, the covert carrier filtering strategy solely determines which packets are chosen as covert carriers. Once the carriers are selected, the processes of data embedding, transmission, and extraction are identical to those in traditional covert channels. Therefore, the introduction of a filtering strategy should not affect the bit error rate (BER) or the effective covert data transmission rate (ETR) under non-ideal network conditions. This experiment aims to validate this theoretical inference.

We employed the built-in Linux tool tc (traffic control) to simulate packet loss, reordering, and delay jitter. The experimental setup involves the transmission of video surveillance traffic, which is used to construct a network timing covert channel. The $base_interval$ was set to $100 ms$. The evaluation metrics selected were the ETR and the BER, defined as follows: 

\begin{equation}
	ETR=\frac{number \ of \  received \ covert \ bits}{total \ number \ of \ covert \ bits}  \times 100\%
\end{equation}
\begin{equation}
	BER=\frac{number \ of \ incorrectly\ received \ covert \ bits}{total \ number \ of \ covert \ bits}  \times 100\%
\end{equation}

\begin{figure}[hptb]
	\centering
	\includegraphics[scale=0.8]{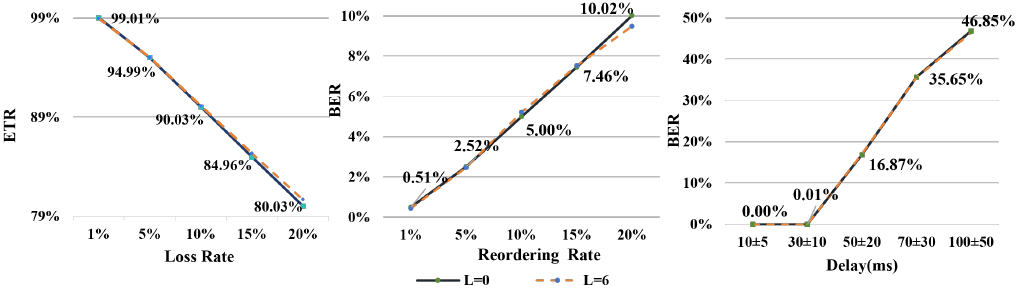}
	\caption{Robustness of Network Timing Covert Channel in Different Network Environment Conditions.}
	\label{Robust_Result}
\end{figure}

The Fig.\ref{Robust_Result} presents the experimental results when $L = 0$ (without filtering strategy) and when $L = 6$. Under all tested conditions, no significant differences are observed in either ETR or BER before and after introducing the filtering strategy. These results indicate that the filtering strategy does not introduce additional robustness losses.

\subsubsection{Covertness Experiments}
To evaluate the effectiveness of the covert carrier filtering strategy in enhancing the covertness of network covert channels, we compared the classification performance of network storage and timing covert channels under different filter key sizes $L$. The classifier’s AUC was used as the primary performance metric.

\paragraph{Network Storage Covert Channel}

\begin{figure}[hptb]
	\centering
	\includegraphics[scale=0.55]{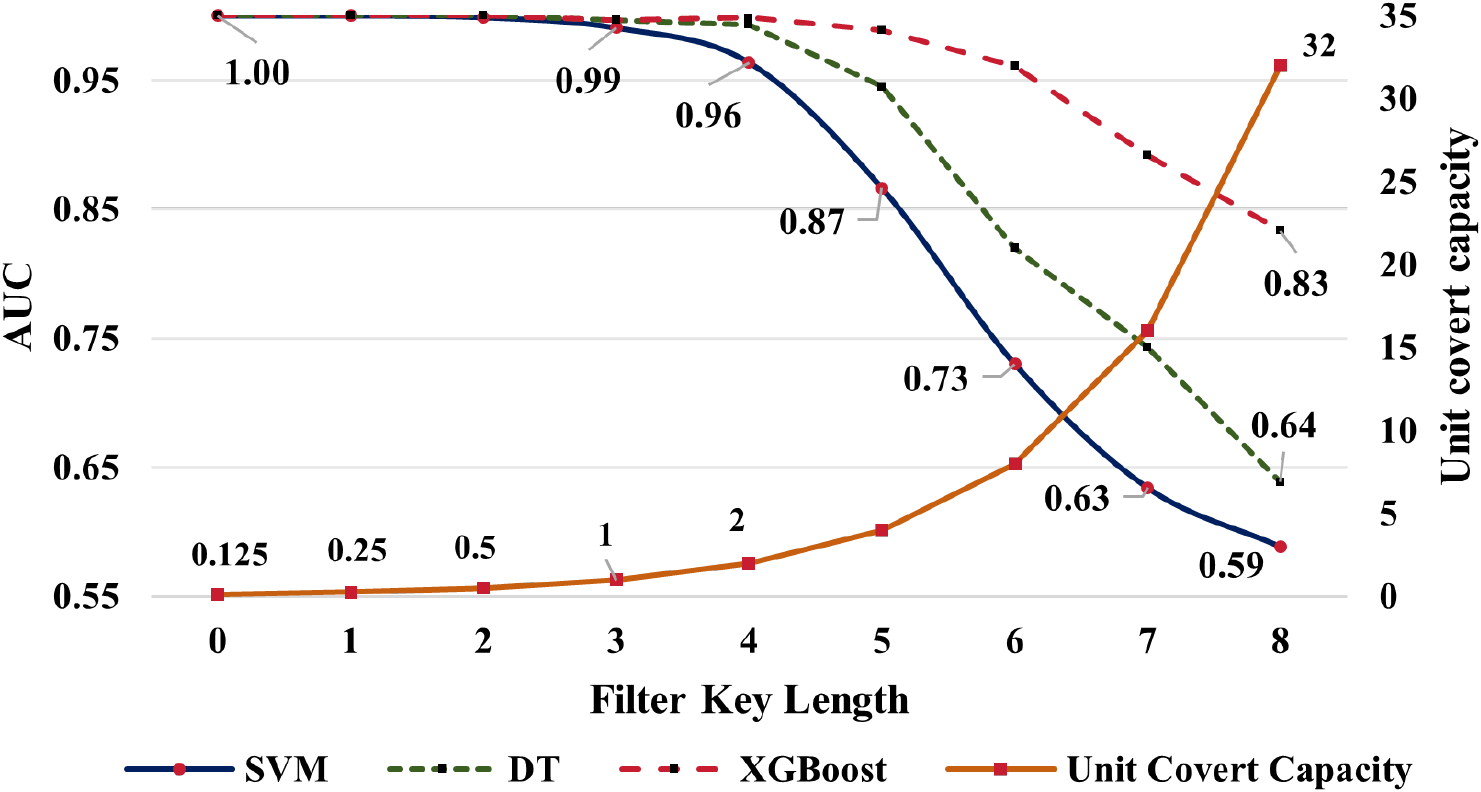}
	\caption{Performance of the network storage covert channel under different filter key sizes.}
	\label{NSCC_Result}
\end{figure}

In this experiment, we extracted the field feature vector $V=(\hat{f}\left(P\right),C_v,H,R\left(\tau\right))$ from the IP protocol’s DS and ID fields, as well as from the TCP protocol’s Seq and Ack fields. Fig.\ref{NSCC_Result} illustrates the classification results of the network storage covert channel under different $L$. As the $L$ increases, the AUC values of the three classifiers show a downward trend.

Additionally, we evaluated the classifier’s baseline performance on normal traffic, which achieved an AUC of 0.51. Without the filtering strategy ($L=0$), the classifier easily learned the stable modification patterns in the DS and ID fields caused by large-scale alterations, achieving a high AUC of 1.0. When a filtering strategy is introduced and $L$ is greater than 6, the AUC values of the three classifiers decrease significantly. For SVM, its classification results approach the level of random guessing.

\paragraph{Network Timing Covert Channel}

\begin{figure}[hptb]
	\centering
	\includegraphics[scale=0.55]{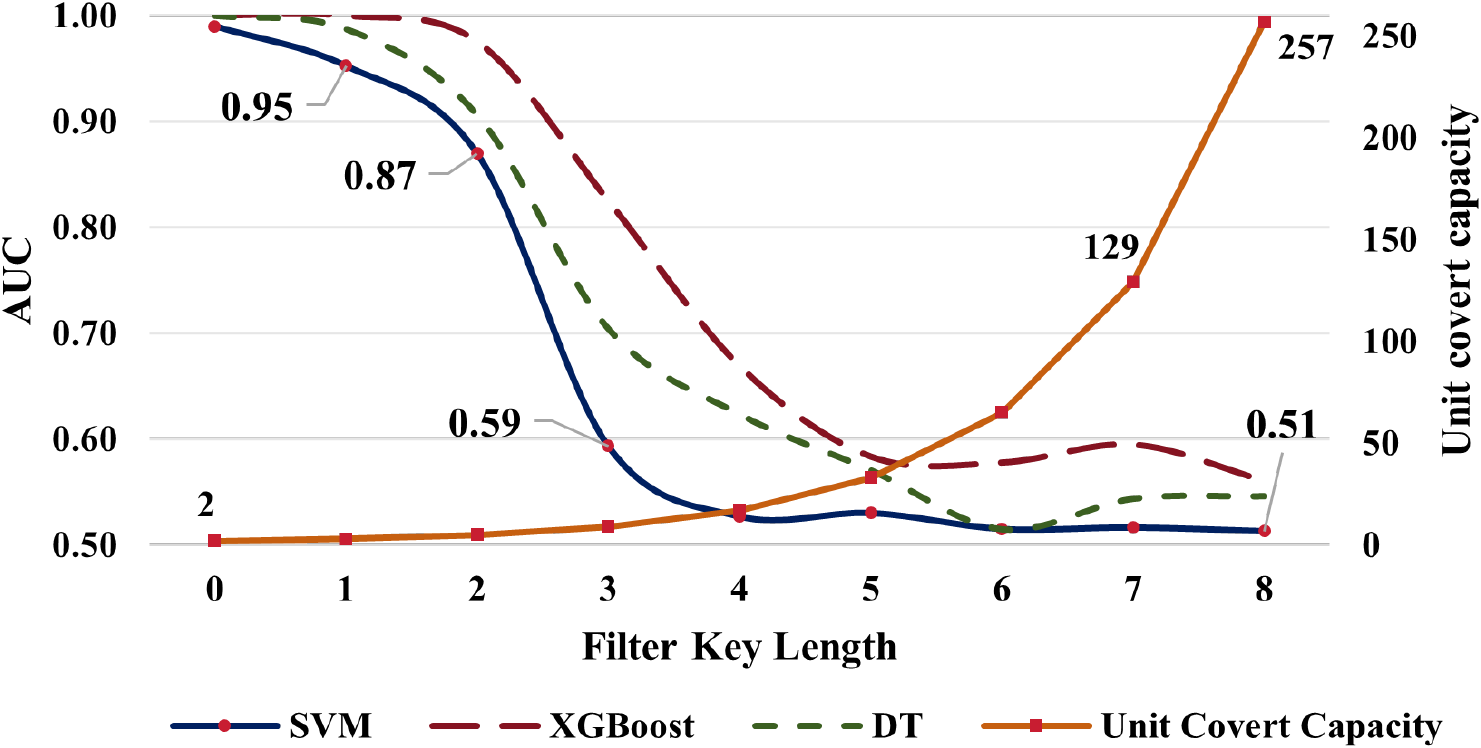}
	\caption{Performance of the network timing covert channel under different filter key sizes.}
	\label{NTCC_Result}
\end{figure}

We extracted inter-packet timing features using a sliding time window of 1,000 packets. The experimental results for the network timing covert channel exhibit a pattern like that of the network storage channel. As shown in Fig.\ref{NTCC_Result}, when the $L$ increases, the temporal regularity of the covert traffic is effectively disrupted. We tested the classification ability of three classifiers for normal traffic, with an average AUC of 0.55. In addition, we separately examined the impact of the time delay introduced by the covert carrier filtering strategy, and test results showed an average AUC of 0.57.

The AUC of a network timing covert channel without the strategy $(L=0)$ is as high as 0.99. When the $L = 5$, the AUC of all three classifiers drops below 0.6. This demonstrates that the strategy effectively conceals the timing features by transforming continuous modulation into sparse, pseudo-random modulation events. Consequently, the resulting timing variations are masked by the network’s inherent jitter, rendering classifiers largely ineffective.

By combining the results from both channel types, we observe that when the $L = 6$, the proposed network covert channels achieve a high level of covertness. Even against advanced machine learning–based classifiers, the covert traffic remains statistically indistinguishable from normal traffic.

\subsubsection{Processing Time Analysis}

Table \ref{Timing} presents the average additional processing time introduced at the CS for performing the filtering and embedding operations on a single packet.

\begin{table}[hptb]
	\centering
	\caption{The average processing time of a single data packet}
	\label{Timing}
	\renewcommand{\arraystretch}{1.2}
	\begin{tabular}{|ccc|c|c|}
		\hline
		\multicolumn{3}{|c|}{Network Covert Channel Types}     & Storage & Timing   \\ \hline
		\multicolumn{3}{|c|}{Average Packet Payload Size}      & 709byte & 1146byte \\ \hline
		\multicolumn{1}{|c|}{\multirow{10}{3cm}{Average processing time per data packet ($\mu s$)}} & \multicolumn{2}{p{4.5cm}|}{covert data embedding without the strategy}                      & 4.1     & 0.78     \\ \cline{2-5} 
		\multicolumn{1}{|c|}{}                                                              & \multicolumn{2}{c|}{the strategy}                                                    & 4.11    & 7.44     \\ \cline{2-5} 
		\multicolumn{1}{|c|}{}                                                              & \multicolumn{1}{c|}{\multirow{8}{3.5cm}{covert data embedding under the strategy}} & $L=1$ & 6.48    & 7.87     \\ \cline{3-5} 
		\multicolumn{1}{|c|}{}                                                              & \multicolumn{1}{c|}{}                                                          & $L=2$ & 6.24    & 7.8      \\ \cline{3-5} 
		\multicolumn{1}{|c|}{}                                                              & \multicolumn{1}{c|}{}                                                          & $L=3$ & 6.02    & 7.73     \\ \cline{3-5} 
		\multicolumn{1}{|c|}{}                                                              & \multicolumn{1}{c|}{}                                                          & $L=4$ & 5.91    & 7.59     \\ \cline{3-5} 
		\multicolumn{1}{|c|}{}                                                              & \multicolumn{1}{c|}{}                                                          & $L=5$ & 5.87    & 7.54     \\ \cline{3-5} 
		\multicolumn{1}{|c|}{}                                                              & \multicolumn{1}{c|}{}                                                          & $L=6$ & 5.78    & 7.63     \\ \cline{3-5} 
		\multicolumn{1}{|c|}{}                                                              & \multicolumn{1}{c|}{}                                                          & $L=7$ & 5.7     & 7.53     \\ \cline{3-5} 
		\multicolumn{1}{|c|}{}                                                              & \multicolumn{1}{c|}{}                                                          & $L=8$ & 5.75    & 7.63     \\ \hline
	\end{tabular}
\end{table}

After introducing the covert carrier filtering strategy into the network covert channel, the packet processing time is primarily dominated by hash computation and filtering decision operations, resulting in additional processing overhead. Depending on the packet payload size, the filtering time for a single packet is maintained within 8 $\mu s$. 


This overhead is relatively low and does not impose a performance bottleneck on the system. It neither causes packet loss due to processing delays nor significantly increases end-to-end latency, demonstrating the feasibility of deploying the proposed strategy in practical network environments. Furthermore, the lightweight nature of the approach can be further enhanced through algorithmic optimization, such as employing lightweight hash algorithms.

\subsubsection{Comparison with Related Work}

Wendzel et al. \cite{paper_12} proposed DYST, a fully passive network covert channel. The core idea of DYST is to separate the control channel from the data channel. Specifically, the CS listens to broadcast traffic within a local area network (which the CR can also receive) and computes a hash over the packet payload to generate a fixed-length sequence. This sequence is then compared with the covert data to be transmitted. When a match is found, the CS notifies the CR through the control channel by generating an ARP broadcast, thereby completing covert data transmission. DYST includes two versions: DYST-Basic and DYST-Ext. The latter functions similarly to DYST-Basic but uses an encoding scheme where the secret message contains only h-c bits, and a checksum of c bits is appended to form an h-bit encoded message.

Essentially, DYST employs a hash function as a filter, selecting suitable packets whose hash values match the covert data currently being transmitted. As shown in Table \ref{Comparsion}, we compare our covert carrier filtering strategy with DYST in terms of the proportion of usable covert carriers and the unit covert capacity. It can be observed that both approaches depend on the number of matching bits required by the hash function. In terms of unit covert capacity, when each packet carries 8 bits of covert data, our proposed filtering strategy outperforms DYST-Basic but is slightly inferior to DYST-Ext. However, when each packet carries more than 16 bits of covert data, our method achieves superior performance compared with both DYST variants. Although the two approaches differ in their underlying mechanisms, they both enhance covert channel covertness by sacrificing channel capacity. Compared to DYST, our method provides greater flexibility through adjustable filter key sizes and periodic key updates, allowing fine-grained control over the trade-off between covertness and capacity.

\begin{table}[!hptb]
	\centering
	\caption{Covert Channel Performance Comparison}
	\renewcommand{\arraystretch}{1.1}
	\label{Comparsion}
	\begin{tabularx}{\textwidth}{>{\centering\arraybackslash}p{4cm}|>{\centering\arraybackslash}p{3cm}|>{\centering\arraybackslash}p{3cm}|>{\centering\arraybackslash}p{2cm}}
		\hline
		\textbf{Types} & \textbf{A single packet hides the number of data bits} & \textbf{Proportion of covert carriers} & \textbf{$U_{cc}$} \\ \hline
		\multirow{4}{*}{DYST-Basic}     & 8bit     & 0.3813\%       & 32.8     \\ \cline{2-4} 
		& 12bit        & 0.2673\%            & 31.2        \\ \cline{2-4} 
		& 16bit                                                  & 0.0022\%                               & 2840              \\ \cline{2-4} 
		& 21bit                                                  & 0.0000\%                               & -                 \\ \hline
		\multirow{4}{*}{DYST-Ext}                                                              & 8bit                                                   & 3.516\%                                & 3.6               \\ \cline{2-4} 
		& 12bit                                                  & 1.9035\%                               & 4.4               \\ \cline{2-4} 
		& 16bit                                                  & 1.0654\%                               & 5.9               \\ \cline{2-4} 
		& 21bit                                                  & 0.3508\%                               & 13.6              \\ \hline
		\multirow{4}{*}{\begin{tabular}[c]{@{}c@{}}Covert Carrier Filter\\ ($L=6$)\end{tabular}} & 8bit    & \multirow{4}{*}{1.56\%}                & 8                 \\ \cline{2-2} \cline{4-4} 
		& 12bit         &         & 5.3      \\ \cline{2-2} \cline{4-4} 
		& 16bit                    &                        & 4                 \\ \cline{2-2} \cline{4-4}
		& 21bit       &        & 3.1         \\ \hline
	\end{tabularx}
\end{table}

\section{Limitations and Discussions}
\label{Limitations_Discussions}

\subsection{Generality}

Our experimental evaluation considers two representative types of network covert channel carriers: office network web-browsing traffic and video surveillance traffic. It is worth emphasizing that we have verified the filtering strategy under non-ideal channel conditions. The experimental results show that, under different network conditions, the filtering strategy has almost no impact on the robustness of the network covert channel. This provides indirect but strong support for the generality of the strategy. As part of future work, we will still plan to systematically evaluate the performance of the filtering strategy under more diverse network protocols and traffic patterns, such as Internet-of-Things traffic and industrial control network traffic.

\subsection{Adaptivity}

Our implementation adopts a static key and a fixed filtering ratio. However, in real-world networks, the statistical characteristics of background traffic and network conditions are inherently dynamic. Fixed parameters may therefore lead to inefficient utilization of covert carriers. A straightforward enhancement is to introduce time-based key updates. For example, the CS\&CR may periodically derive new filtering keys from a pre-shared master key. In conjunction with network conditions, the filtering ratio can be moderately adjusted within predefined security bounds to achieve basic adaptivity. A more advanced approach is to construct an adaptive framework based on artificial intelligence. By continuously extracting statistical features of background traffic and applying reinforcement learning models, system parameters can be dynamically optimized. The objective is to maximize effective throughput under given covertness constraints, thereby enabling intelligent and continuous optimization of the covertness–capacity trade-off in dynamic network environments.

\section{Conclusion and Future Work}
\label{Conclusion}

This paper addresses a core limitation of traditional network covert channels: their susceptibility to detection due to algorithmic dependence. We propose a covert carrier filtering strategy based on Hash. By employing cryptographically secure hash functions in conjunction with a shared key, the strategy dynamically selects a pseudorandom subset of packets from network traffic as covert carriers, thereby shifting the covertness of the network covert channel from algorithmic secrecy to key-based security. We analyze the covertness of the proposed covert carrier filtering strategy by constructing a network covert channel model, and we design and implement a covert carrier filtering strategy based on the SHA-256 Hash. The strategy is validated and experimentally evaluated in two typical network covert channel scenarios. Experimental results demonstrate that the strategy can enhance the detection resistance of network covert channels. In summary, the covert carrier filtering strategy offers an effective new paradigm for enhancing the survivability of network covert channels in adversarial environments. By addressing existing limitations such as static parameter configurations and insufficient robustness, network covert channels can evolve toward greater intelligence and resilience.

\bibliographystyle{elsarticle-num-names}
\bibliography{reference}

\end{document}